\documentclass[manuscript]{aastex}

\def \msun   {\hbox{M$_\odot$}}

\def \lsun     {\hbox{L$_{\odot}$}}

\shorttitle{Masses and Distance of NTTS 045251$+$3016}
\shortauthors{A,B,C et al.}

\begin{document}

\title{Masses and Distance of the Young Binary NTTS 045251$+$3016}

\author{M. Simon\altaffilmark{1},  G. H. Schaefer\altaffilmark{2}, 
L. Prato\altaffilmark{3}, Dary Ru\'{i}z-Rodr\'{i}guez\altaffilmark{3},
N. Karnath\altaffilmark{3},  O. G. Franz\altaffilmark{3},
and L. H. Wasserman\altaffilmark{3}}

\altaffiltext{1}{Department of Physics and Astronomy, Stony Brook University,
    Stony Brook, NY 11794-3800, USA  michal.simon@stonybrook.edu}
\altaffiltext{2}{The CHARA Array of Georgia State University, Mount Wilson
Observatory, Mount Wilson, CA 91023, USA}
\altaffiltext{3}{Lowell Observatory, 1400 West Mars Hill Rd., Flagstaff, AZ
 86001, USA}

\begin{abstract}

As part of our continuing campaign to measure the masses of 
pre-main sequence (PMS) stars 
dynamically and thus to assess the reliability of the discrepant theoretical 
calculations of contraction to the main sequence, we present new results 
for NTTS 045251$+$3016, a visual and double-lined spectroscopic binary
in the Taurus-Star Forming Region (SFR).
We obtained new high angular resolution astrometry
and high spectral resolution spectroscopy at Keck Observatory.
The new data lead to a significant revision of previously published orbital
parameters.  In particular,  we find that the masses of the primary 
and secondary are $0.86 \pm 0.11$~and $0.55 \pm 0.05$ \msun,
respectively, smaller than previously reported,  and that the system 
lies $158.7 \pm 3.9$ pc from the sun, further than previously reported.

\end{abstract}

\keywords{binaries: visual -stars: Pre-Main Sequence---- stars: 
fundamental parameters----stars: individual (NTTS 045251$+$3016)}

\section{Introduction}

The  mass and age of a  pre-main sequence (PMS) star 
are usually estimated by its location in the H-R diagram 
(HRD) relative to theoretical calculations of PMS evolution.
Unfortunately, for stars less massive than $\sim1.5$~\msun,
these estimates vary depending on the calculations used 
(e.g., Simon 2008; Hillenbrand \& White 2004).  For example,  for 
an M0 spectral type star younger than $\sim10$ Myr with 
luminosity L=0.5 \lsun, the tracks can yield masses  
discrepant by a few tenths of a solar mass and 
ages discrepant by factors of 2 to 3.
Dynamical techniques provide the only absolutely trustworthy
measurements of mass and thus the only means to connect  the
theoretical calculations  with reliable empirical input.  We are
pursuing a program to measure PMS stellar masses dynamically
(e.g., Schaefer et al.\ 2012 and references therein) with
sufficient precision and accuracy to identify a reliable set of
evolutionary tracks for observers and
to identify improvements that may be necessary to implement for theorists.

Walter et al. (1988) identified NTTS 045251$+$3016  (V397 Aur, TAP 57, HBC 427)
as a single-lined pre-main sequence (PMS) binary in the 
Taurus-Auriga star forming region (SFR) on the basis of its visible light spectroscopy
and X-ray emission.  Steffen et al. (2001) 
resolved the binary using the {\it Fine Guidance
Sensors} (FGS) of the {\it Hubble Space Telescope} and combined the results with visible 
light spectroscopy spanning 14 years and infrared (IR) spectroscopic 
measurements that detected the secondary.  They derived the orbital
parameters of the system as a visual binary (VB)
and double-lined spectroscopic binary (SB2).  In particular,
Steffen et al.\ reported  masses of the components, orbital period, 
semi-major axis, inclination, and  the distance to the binary.
These parameters  were derived from
astrometry covering  about half the 6.9 year orbit, single-lined radial 
velocities (RVs) of the primary covering the full period, and only two IR 
spectroscopic measurements of both component RVs close together in 
phase but fortuitously at maximum RV
separation.   As a result, the precisions of the primary and 
secondary masses are $13\%$ and $ 11\%$, respectively, 
insufficient to discriminate definitively among theoretical
calculations of PMS evolution (e.g., Simon et al. 2000).   
We therefore  included NTTS 045251$+$3016 in our current program.
 
It is now possible to achieve angular resolution
from the ground that is comparable to or exceeds
that of the {\it FGS}.  We present such results here.
\S 2 describes our new observations
obtained using adaptive optics (AO),  non-redundant masking (NRM),
two-telescope interferometry (archival data), 
and high resolution spectroscopy, all in the near-IR. 
To our surprise, we found that the new astrometry is inconsistent
with Steffen et al.'s. \S 3 discusses this discrepancy, presents
revised orbital parameters of the binary including new values 
for the component masses and distance, and
compares the masses with  theoretical
calculations of PMS evolution.  \S 4 summarizes our
results.

\section{ Observations and Data Reduction}
 
All of our observations (Table 1) were obtained with facility
instrumentation at the two telescopes of the W.M. Keck Observatory. 
We describe the observations and their reduction  in this section.

\subsection{Two-telescope Interferometry}

NTTS 045251$+$3016 was observed on seven nights with the two 
telescopes of the Keck Observatory operating as an IR interferometer
(Colavita et al. 2004; Wizinowich et al. 2004).
This mode was decommissioned in 2012.  All Keck Interferometer (KI) data 
are now in the public domain; we obtained the data for NTTS 045251$+$3016
from the archives maintained at the {\it NASA Exoplanet Science Institute} (NExScI)
at the California Institute of Technology.  The observations were originally 
proposed and planned by A.\ Sargent and A.\ Boden and were carried out by
the KI observing staff.

Operating in its visibility-amplitude mode, the KI measured the visibility,
or fringe contrast, of the target.  The normalized visibility, $V^2$,
of a binary when the diameter of each component is unresolved, a
safe assumption for stars at the 140 pc distance of the Taurus SFR (Kenyon et al. 1994), is
\begin{equation}
V^2 = {1 + r^2 + 2r\cos{(2\pi~{\bf B\cdot s} /\lambda)}\over
(1+r)^2}
\end{equation}
where $\lambda$~is the wavelength of observation, and $r$, the flux ratio 
of the components (e.g., Boden 2000; Berger \& Segransan 2007). The projection of the binary separation  
vector ${\bf s}$ on the interferometer baseline ${\bf B}$ is given
by  (${\bf B\cdot s}$).  The 85-m baseline of the KI is 
oriented $38^\circ$ east of north  (Colavita et al. 2004). 
The components of $({\bf B}/\lambda)$ projected on the sky are the spatial 
frequencies $u$ and $v$ commonly used in interferometry.  The Earth's rotation
presents continually varying projections of the baseline with respect
to the target and the interferometer measures $V^2$~ varying with
$u$ and $v$, or  equivalently, time.  Analysis of the measured $V^2(u,v)$ is
accomplished by fitting a binary model to the data.  

The observations of NTTS 045251$+$3016 were interspersed with those of unresolved
calibrators; Table 2 lists the calibrator spectral types, V and K-band 
magnitudes, and adopted angular diameters.  The diameters were obtained from 
the SearchCal software tool developed by the JMMC Working Group (Bonneau 
et al.\ 2006, 2011).  The visibilities were calibrated using the wbcalib 
program distributed by NExScI.  The uncertainties are estimated from the 
scatter in the interferometric scans obtained on the target and calibrators.  
When running wbcalib, we applied the ratio correction, to account for when 
the flux from the two telescopes of the interferometer is unbalanced, and the 
flux bias correction, to account for the visibility dependence on target 
brightness.  We rejected visibility measurements when the ratio correction 
exceeded 1.5.  Table 3 lists the time of observations, central wavelength,  
calibrated visibilities, and their corresponding $u$ and $v$ coordinates.

Given the limited ($u,v$) coverage on the sky, we fit the orbit (see
Sect. 3.2) directly to the visibilities rather than measuring the
separation and position angle during each epoch separately.  This
allows us to use additional dynamical information from the
spectroscopic,  AO, and NRM measurements to help constrain the binary 
position during the KI epochs.  This is critical for nights with only
a single KI observation of NTTS 045251$+$3016 and also useful for nights 
with multiple KI observations but with limited rotation of the baseline on the sky.  

We encountered a few difficulties when fitting the KI visibilities.
First, we were not able to get a reasonable fit for the data on UT
2006 Dec 08.  Inspection of the observing log from that night revealed
that the counts from both telescopes were lower than expected,
windshake on Keck II was $\sim$ 10 times higher than on Keck I when
pointing into the wind, and significant aberrations were present in the
images obtained from the Keck II angle trackers.  Additionally, the 
ratio correction between the two telescopes was much higher for 
observations of NTTS 045251$+$3016 compared with other targets and 
calibrators during the night, and varied over the
sequence on NTTS 045251$+$3016.  Because of these issues, we opted not 
to include the KI data from 2006 Dec 08 in the fit.

The second complication was during the KI epochs when the binary was
widely separated.  In 2002 and 2004, the components in NTTS
045251$+$3016 were separated by 35 mas and 29 mas, respectively (based 
on the orbital parameters presented in Sect 3.2).  With a K-band
filter width of 0.3~$\mu$m, the coherence length
($\frac{\lambda^2}{\Delta\lambda}$) of the KI corresponds to
$\sim$~38~mas on the sky.  Therefore, the width of the fringe envelope 
from each component in the binary becomes important, so we added a
bandwidth smearing term to the visibilities (e.g., Bridle \& Schwab
1999; Lachaume \& Berger 2012).  However, these epochs were
complicated by one additional factor.  The field of view of the KI is
set by the width of the single mode fibers used in the fringe tracker
and corresponds to $\sim$~45~mas in the K-band.  Therefore, at wide 
separations, not all of the light from both components will get
through.  Using our AO images of NTTS 045251$+$3016 taken in the K-band 
on 2009 Oct 25, when the binary is separated by $\sim$~36~mas
(\S 2.2),  we estimate that if the KI field of view is centered on
the center of light, then we expect to lose 15\% more light from the
secondary as compared with the primary in an aperture of 45~mas.
Additionally, the amount of flux lost from the companion would depend 
on the seeing and quality of the AO correction; this would effectively
change the flux ratio between the components on those nights.  If 
there were enough KI measurements on these nights to sample the binary 
visibility curve sufficiently, then we would be able to measure the
effective flux ratio independently for each night.  However, because
there was only one observing block on NTTS 045251$+$3016 during each 
of the nights when the binary was widely separated, we opted not to
use the KI data from 2002$-$2004 in our orbital fit because of our
inability to determine the arbitrary scaling of the visibilities.

The KI visibilities included in the orbit fit (\S 3.2) are marked
by an asterisk in the last column of Table 3.  Figure 1 plots the
$V^2$ values for each of these nights.  The largest number of KI
measurements was obtained on 2005 Oct 25; this night samples the 
binary visibility curve well and adds valuable dynamical
constraints on the orbit fit.

\subsection{Adaptive Optics Imaging}

We used NIRC2, the facility near IR camera  (Wizinowich et al.\ 
2000), to obtain adaptive optics (AO) images on four occasions (Table 1).  
The images were recorded in the narrow-field mode with a plate scale  
$9.952\pm 0.002$ mas per pixel and an orientation of $0.252^\circ \pm 0.009^\circ$ east 
of north (Yelda et al. 2010).  The PMS stars DN Tau and LkCa~19 served as 
point spread function (PSF) references; their separations from 
NTTS 045251$+$3016 are  $ 7.7^\circ$ and $5.4^\circ$, respectively.  
Neither PSF star is known to have companions (Leinert et al. 1993; Prato et al. 2008;
Kraus et al. 2011) and, at the  distance of the Taurus SFR, their diameters are 
unresolvable by the AO observations.

Table 4 gives AO imaging specifications of the NTTS 045251$+$3016 observations.
The target was observed using a standard three-point image dither pattern with 
$\sim2''$ offset.  Each image was comprised 
of 10 coadded frames, each frame with an exposure time of 0.18$-$1.5 s.  
The PSF reference star was observed immediately before or after the target
using the same dither pattern and AO frame rate to minimize changes 
in the shape and structure of the PSF.  We collected a total of 6$-$12 
images on target and 3$-$12 images of the PSF calibrator during each night.
Post-observing processing of the raw
images, and subsequent analysis were the same as we described in
Schaefer et al.\ (2012).  The images were flatfielded using dark-subtracted, 
median-filtered dome flats.  Pairs of dithered exposures were subtracted to 
remove the sky background.  

Figure 2 shows the co-added images of NTTS 045251$+$3016 and the PSF reference
stars.  The NTTS 045251$+$3016
binary was reliably resolved in all the observations except on UT 2011
Oct 12.  Observations using the non-redundant masking technique (next
section) did resolve the binary on that date at a separation of $\sim16$ mas.  
Fitting the standard AO data from the same night we measured a separation of 
$\sim17$ mas at a consistent position angle; however, the uncertainties were 
about four times larger compared with the results obtained with the aperture mask.
The Airy criterion for the diffraction limited resolution of a 10-m diameter 
telescope is 40 mas at 1.6 $\mu$m.  We are not confident that we can measure 
separations half this size reliably using conventional AO observations without 
a simultaneous PSF reference in the field of view.  For observations that
require a separate PSF reference, as in the case of NTTS 045251$+$3016, the
time-variability of the AO correction limits our ability to resolve binaries 
with separations significantly smaller than the diffraction limit.

For each night, we used the PSF reference star to construct models of the 
binary while searching through a grid of separations and flux ratios and 
selecting the solution where the $\chi^2$ between the data and model reached 
a minimum.  Uncertainties were determined by analyzing multiple images 
individually and computing the standard deviation.  We applied the
geometric distortion solution of Yelda et al. (2010). 
Table 5 lists the separation $\rho$, position angle $PA$ measured east of 
north, and flux ratio $r$  derived from our NIRC2
observations.  

\subsection{Non-Redundant Masking  Interferometry without Tears}

NIRC2 includes a 9-hole mask that can be positioned in the pupil
plane for observations by non-redundant masking interferometry (NRM)\footnote{
The technique is also known as sparse aperture imaging. P. Tuthill's 
layout for the apertures is given at the NIRC2 website:
www2.keck.hawaii.edu/inst/nirc2/specs.html}.   The 9 apertures provide
36 independent and unique baselines.   Two apertures at the ends of a
baseline form a simple two-element 
interferometer and produce an interferogram on the detector array with
fringe spacing $\lambda/B$ where $B$ designates the baseline 
length as in eqn. (1).   According to the usual criterion that two
objects are resolvable when the maximum of the fringe pattern of one falls 
on the first minimum of the other, a two-element interferometer 
achieves resolution $\sim0.5 \lambda/B$,  better than twice that of 
a uniformly illuminated aperture of diameter B.  The Fourier transform
of the interferogram produces signal at two locations in the $u,v$
plane because the two apertures defining the baseline  are
indistinguishable.  In actuality, the signal is spread
over a small patch in the $u,v$ plane, the result of convolution with
the finite aperture size.  Lloyd et al. (2006) call these distributions of
signal ``splodges'' (see their Figure 3); the term is
so evocative and  apt that we adopt it here.
The amplitude squared of this Fourier transform, the
power spectrum of the splodges, is the $V^2$ of eqn (1).  
The phase, defined as  $\tan^{-1} (Im/Re)$, where $Im$ and
$Re$  are the imaginary and real parts of the Fourier transform,
completes the structural information of the target.   In contrast to
the KI which may be considered a two-aperture synthesis interferometer,
 an NRM-equipped camera when operating on an equatorial mount telescope produces
a stationary signal pattern in the $u,v$ plane. Because Keck II is an elevation-azimuth
mounted telescope the pattern produced on NIRC2's detector array
rotates during the observation.

The 36 baselines produce 36 interferograms superposed on the detector
array.  The two-dimensional Fourier transform of the signal recorded on
the array produces 72 splodges in the $u,v$ plane.  The phase of the 
2-D transform not only carries the structural information of the
target, but also includes unavoidable  phase errors 
induced by the atmosphere,  incomplete wave-front correction, and 
systemics in the camera and  processing.  These errors cancel in
forming the closure phase,  the sum of the splodge phases around a 
triangle formed by 3 apertures
(Baldwin et al. 1986; Nakajima et al. 1989; Tuthill et al. 2000).  
The splodges produced by the 9 apertures yield 84 closure
phases, the number of combinations of 9 apertures taken 3 at a time. 
Closure phases measured on the target are calibrated by observations
of a nearby unresolved star.  When the target is a simple binary, the 
calibrated closure phases are then fit to a model to determine its 
$\rho$, PA, and $r$ (e.g., Lloyd et al. 2006; Martinache et al. 2009; 
Kraus et al. 2011.)

We made all the NRM observations in the H-band, taking
a series of  dithered integrations on NTTS 045251$+$3016 
followed by a series on LkCa 19.  The numbers of integrations
in each observation are given in Table 4. After dark-subtracting and
flat-fielding the data as for other NIRC2 images, we calculated
the closure phases for target and calibrator using procedures written
at Stony Brook.   Differencing
these yielded calibrated closure phases for NTTS 045251$+$3016.
We determined the best fitting binary parameters by minimizing the
$\chi^2$ of a model with respect to the data.  In actual practice, the 
modeling routine requires two instrumental parameters:  the rotation 
of the mask with respect to the detector array of the camera and 
the scale of the calculated Fourier transform of the interferogram 
with respect to the observed splodges.   We determined these by 
identifying the correspondence of splodge locations of the calibrator 
with the Fourier transform of the interferogram of the unrotated
mask.  We checked the image scale and rotation by observing DF Tau, a
$\sim100$ mas separation PMS binary, with both AO imaging and NRM on 
UT 2013 Jan 27; these results will be reported separately (Schaefer et
al. 2013, in prep).
The analysis also requires a telescope parameter, the
orientation of the camera with respect to the sky.  This changes
as the altitude-azimuth mounted telescope tracks a star.  It 
is calculated by the Keck facility software and provided in the image
headers. 

We analyzed the integrations of each observing
session in groups of eight. Over the total time required for each 
group, the sky rotation was smaller  than $1.6^\circ$.
For each of the UT 2011 Jan 24 (Figure 3, {\it top}) 
and 2011 Oct 12 (Figure 3, {\it bottom}) sessions, we show the averaged
closure phases for the first 8 integrations as a function of the
longest baseline in the triangle for which a closure phase is measured. 
The best fit parameters in Table 5 are for the full data sets
and the uncertainties are calculated from the analyses of the four
groups of  8 integrations. The NRM-derived results for UT 2011 Jan 24 are in good
agreement with those obtained with conventional AO-assisted imaging.
On UT 2011 Oct 12, NRM resolved the binary components at separation $\sim16$ mas,
demonstrating that it is possible to reach angular resolution as small as
$40\%$ of that given by the Airy criterion.  If the binary were unresolved,
all the closure phases would be zero.  That the closure phases
are smaller on the longest baselines in the October, 2011 observation than
in the January, 2011 observation is a result of the smaller separation
in October.  At secondary/primary
 flux ratio $\sim0.4$ in the H-band, our observations did not 
stress the contrast limits of the NRM technique; however, Kraus et al. (2011)
showed that it is possible to reach  15:1 contrast in K-band observations of a 
large sample of stars in the Taurus-Auriga SFR.

Table 5 also includes our best fit parameters to the UT 2008 Dec 23 NRM
data obtained and analyzed by Kraus et al. (2011)\footnote{We obtained 
these data from the archives maintained at the {\it NASA Exoplanet 
Science Institute} (NExScI).}.  We analyzed the data independently
because we wanted another check on our analysis procedures and
because we wanted to use parameters derived uniformly
for the orbital fit (\S 3).  The separation, position angle and
component flux ratio we derived for this observation  are entirely 
consistent with  Kraus et al.'s values  except
that the position angle uncertainty is larger.

\subsection{High Resolution IR Spectroscopy}

We measured the RVs, rotational broadening, and
determined the best matching spectral type for the primary and
secondary using high-resolution, near-IR spectra taken with
NIRSPEC on the Keck II telescope (McLean et al. 2000).  Our observations
were centered at $\sim1.555\mu$m (order 49); the $0.''288$ (2-pixel) wide slit provided 
spectral resolution 30,000.   These observations did not use AO.
As the component spectra were angularly unresolved,
we used the 2-D cross-correlation procedure
developed at Lowell Observatory (e.g., Mace et al. 2012)
to recover the RVs of the two binary components.
The best solution for the spectra was found with a K5V primary 
(HR 8085) and an M0V secondary (GL 763), corresponding to T$_{\rm eff}=4415$ K 
and 3845 K (Luhman et al. 2003), respectively, each rotationally 
broadened to 13 km s$^{-1}$.  The RV measurements 
are listed in Table 6.   In Figure 4 these values are shown together 
with the visible light and IR measurements previously 
reported by Steffen et al. (2001).

\section{Results and Discussion}

\subsection{Spatially Resolved Binary Separations}

Figure 5 shows our AO and NRM astrometric measurements and the computed 
location of the binary during the times of the KI observations; the orbit was fit directly
to the KI visibilities (\S 2.1).  The parameters of our orbital fit
are reported in Table 7 and discussed in the following section.
Figure 5 also shows the earlier FGS measurements of Steffen et
al. (2001); clearly their  and our new measurements are discrepant.
We suspect a fault in the FGS results because the new measurements 
appear consistent with each other and were derived by three different techniques.  

The FGS is sensitive in the visible and operates as a scanning
interferometer along two orthogonal axes.  Using the separation and position
angle measurements reported by Steffen et al.\ (2001) and the spacecraft roll 
angle during each observation, we derived the separations along the FGS $(X,Y)$
axes that they must have measured.  Of the 28 individual 
$(X,Y)$ values for their 14 reported positions, 18 are smaller than 20 mas, 
with 5 under 10 mas.  It has been our experience that it is not possible to 
get a meaningful fit to an FGS transfer function with $\Delta V \gtrsim 2$ mag 
(as reported by Steffen et al.) at such small separations.

We downloaded the FGS data on NTTS 045251$+$3016 from the {\it HST} archive 
and attempted to fit the scans with two different templates, SAO 185689 
and HD 233877.  The magnitude differences and separations we derived varied 
widely for the epochs with small separations.  Interestingly,
we obtained similar fits along the $Y$ axis for the epochs with larger 
separations (30$-$40 mas).  Based on these results, we suspect that the
source of the discrepancy in the astrometric measurements lies in the 
difficulty of detecting a faint component ($\Delta V \gtrsim 2$ mag) 
when the separation is small along one of the axes of the FGS.

\subsection{Orbital Parameters, Masses, and Distance}

The full ensemble of RV data (Figure 4) uniquely determines
$K_1$ and $K_2$, the velocity semi-amplitudes of
the primary and secondary, and hence their mass ratio. The visual orbit
(Figure 5) uniquely
determines the apparent semi-major axis $a$, in angular measure,
and the orbital inclination $i$.  Either the orbital RVs or 
the visual orbit,
or both together, determine the orbital period  $P$, time of
periastron passage $T$, eccentricity $e$, and longitude of the
periastron,  $\omega$.  Together, the visual and spectroscopic orbits
determine the masses of the primary and secondary, $M_1$ and $M_2$,
semi-major axis, $a$,  and the distance to NTTS 045251$+$3016
(e.g., Schaefer et al. 2008):
\begin{equation}
M_1 = {{1.036\times10^{-7}(K_1 +K_2)K_2 P(1-e^2)^{{3\over2}}} \over
  \sin^3{i}}
\end{equation}
and likewise for $M_2$, with $K_1$ replacing the second $K_2$.  Also,
\begin{equation}
a\sin{i} = 0.01375(K_1 +K_2)P(1-e^2)^{{1\over2}} ~~{\rm and}
\end{equation}
\begin{equation}
d= { a({\rm  AU}) \over a('')},
\end{equation}
where the K's are in km/s, P in days, M in solar units, a in $10^6$
km (except in Eqn. 4, where $a$ is given in AU and arcsec), and $d$ is in pc.

We carried out a simultaneous orbit fit to the AO and NRM 
astrometric measurements, the interferometric visibilities, and 
the RVs using a Newton-Raphson technique to determine all 
ten orbital parameters including the center-of-mass velocity $\gamma$, 
and position angle of the nodes, $\Omega$.  We also measured the K-band 
flux ratio from the KI visibilities during the fit.
These values are given in Table 7. 

To demonstrate how the different data sets help determine the orbit, we 
performed a $\chi^2$ search for orbital solutions within the 1\,$\sigma$ 
($\Delta \chi^2 = 1$) and 3\,$\sigma$ ($\Delta \chi^2 = 9$) confidence 
intervals.  We randomly chose values for $a$, $i$, and $\Omega$ and optimized the 
fit for $P$, $T$, $e$, $\omega$, $K_1$, $K_2$, $\gamma$, and the flux 
ratio at K.  Except for the flux
ratio, the latter parameters are well-constrained by the RV measurements while the former 
are derived from the AO, NRM, and KI measurements.  The 1\,$\sigma$ intervals 
obtained for each parameter from the $\chi^2$ search agree well with 
the formal uncertainties determined from the covariance matrix (Table 7).  
In Figure~1, we show how the fit to the KI visibilities changes if we vary
$a, i,$ and $\Omega$ by their 1 and 3\,$\sigma$ uncertainties while
optimizing the remaining parameters.  The same range of orbits is plotted in
the right panel of Figure~5 to show how they fit the AO and NRM
measurements.  Each orbit found through the $\chi^2$ search gives a slightly
different position of the companion during the times of the KI epochs.  In
the left panel of Figure~5 we overplot the $\Delta \chi^2 = 1$ surfaces that
define the 1\,$\sigma$ uncertainty ranges in $\Delta$RA and $\Delta$DEC
(small blue ellipses inside the open circles).

Of the new parameters derived in Table 7, those
determined by the orbital RVs are in reasonable agreement
with the values reported by Steffen et al. (2001).  Our addition of 8
new measurements of the secondary's RV improves the reliability of
its semi-amplitude, $K_2$, and hence of the mass ratio, $q$.  Our new
astrometric orbit affects the semi-major axis in angular measure,
$a''$, and inclination, $i$, and hence the derived masses and
distances (e.g., eqns. 3-4).  Compared to Steffen et al.'s values,
$a('')$ decreases from $\sim32$ to $26$ mas, and $i$ changes 
from $\sim114^\circ$ to $\sim78^\circ$.    As a result,
 the masses of the components decrease, $M_1$ from 1.45 \msun~to
0.86 \msun~and $M_2$ from 0.81 \msun~to 0.55 \msun.  
The precisions of the new mass determinations, $ 0.86\pm0.11$ ~and
$0.55\pm 0.05$~\msun~are hardly improved from the precisions
of Steffen et al.'s values, mostly because we could not extract meaningful results
from the FGS data.  We  compare the new mass values with theoretical 
calculations of PMS calculations in the following section.

The new values of semi-major axis and inclination increase the
distance of NTTS 045251$+$3016 to $\sim158.7 \pm 3.9$~pc with respect to
Steffen et al.'s value, $\sim144.8 \pm 8.3$~pc.  
NTTS+045251$+$3016 thus lies on the far-side of the SFR with
respect to a mean distance estimate of 140 pc (Kenyon et al. 1994).
This is consistent with the spread of distances determined
using the VLBA (Torres et al. 2009), but is inconsistent with Bertout \& Genova's (2006) estimate,
$116^{+14}_{-12}$, probably because their determination used an early
value of the center-of-mass velocity quoted from Walter et al. (1988)  in 
the Herbig and Bell Catalog (Herbig \& Bell 1988).

\subsection{Comparison with Evolutionary Tracks}

We compare the measured masses with the evolutionary tracks of Baraffe 
et al. (1998, BCAH), Siess et al. (2000, SDF), Tognelli 
et al. (2011, Pisa) and the Dartmouth group (Feiden et al. 2011 and
references therein; Dartmouth). To place the 
components of NTTS 045251$+$3016 on Hertzsprung-Russell diagrams (HRDs) 
requires their effective temperatures, $T_{\rm eff}$, and luminosities,
$ L$.  For the primary, we use $T_{\rm eff} = 4345^{+194}_{-138}$ K
derived by Steffen et al. (2001) from consideration of  visible light spectra.
This is within 70 K of the temperature corresponding to the
K5 spectral type determined on the basis of the IR
spectra and converted to $T_{\rm eff}$ using the results in Luhman et al.
(2003; \S 2.4).   We use Steffen et al.'s $T_{\rm eff}$ because the uncertainties are
smaller and because the 70 K difference in results between the two is negligible.
For the secondary we estimated $T_{\rm eff}=3845^{+200}_{-130}$ (Luhman et al.),
corresponding to its M0 spectral type and $\pm$ one subclass uncertainty.

For the BCAH, SDF, and Dartmouth tracks it is convenient to plot the 
absolute magnitudes of the components at H or K
as proxies for the luminosity.  To calculate these we used the 
total apparent 2-MASS magnitudes at H and K, apportioned the
flux to the primary and secondary according to the average of the
flux ratios given in Table 5, and applied the distance modulus
$6.00 \pm 0.05$ mag corresponding to the distance $158.7 \pm 3.9$
pc (Table 7).   We applied no correction for extinction because  
Walter et al.'s (1988) photometry of  NTTS 045251$+$3016
indicates that its extinction at K is negligible.  
The  Pisa tracks are presented in terms of luminosity.  
We calculated luminosities of the components from their $M_K$
values by applying the bolometric correction
and (V-K) color appropriate to their $T_{\rm eff}$ as given by Kenyon
\& Hartmann (1995).   Component luminosities using their $M_H$ values are
consistent.  The uncertainties of the absolute magnitudes and
luminosity include uncertainties in photometry, average flux ratios,
and distance propagated in quadrature.  

Figure 6 shows the primary and secondary plotted on the BCAH, SDF,
Pisa, and Dartmouth evolutionary tracks for ages between 1 and 10
Myr.  We plot the BCAH tracks for mixing length parameter
$\alpha = 1.0$ below 0.6 \msun, and $\alpha=1.9$ for greater
masses.  The SDF tracks use $\alpha=1.605$ and    
the Pisa tracks are available for two values of $\alpha$,
1.20 and 1.68; we show both. The Dartmouth tracks use $\alpha=1.83$
(Dotter et al. 2008).  We consider two questions:\\

\parindent=0.0in

1)  Are the evolutionary tracks consistent with the measured 
masses?\\

2)  What is the age of NTTS 045251$+$3016?\\

The answers are mostly limited by the uncertainties of the primary 
and secondary's masses and $T_{\rm eff}$.  The effects of uncertainties
in luminosity or absolute magnitude are smaller because the
evolutionary tracks are nearly parallel to the abscissa at the ages
we consider.  Table~8 summarizes the range of masses and ages 
derived from the tracks.

\parindent=0.5in

We consider first the primary with measured mass $0.86 \pm 0.11$ \msun.
On the HRD for the Pisa models (Figure 6c) the primary star falls close to the
1.0 \msun~evolutionary track for the $\alpha = 1.68$ model and is
consistent with lower mass tracks corresponding to its measured dynamical mass.
The primary's location on the BCAH, SDF, and Dartmouth tracks is consistent
to a similar extent with its measured mass
as well.  The evolutionary codes adjust the mixing length parameter
to fit the parameters of the Sun.  Evidently this yields 
models that satisfy PMS stars of mass close to 1 \msun.  On all the
HRDs, the $T_{\rm eff}$ uncertainty spans about $\pm 0.1$ \msun, about the
same as the uncertainty in the mass. Improvement in both uncertainties
to at least half the present values would allow a more discriminating
assessment of agreement with the tracks.  The age of the primary is within
$3 \pm 2 $ Myr on all the HRDs.

The secondary star, with mass $0.55 \pm 0.05 $ \msun, is close to agreement with the
Pisa tracks for the $\alpha =1.20 $ models and similarly with the BCAH,
SDF, and Dartmouth models.  The effective temperature is the biggest
limitation preventing a closer assessment of agreement with the tracks.
There is a considerable spread in ages indicated, $\sim3$ to nearly
10 Myr.  There is no reason to suppose that the components are not
coeval.  Improvement in the $T_{\rm eff}$ estimate is necessary to resolve
this discrepancy.  The age of the secondary star is consistent with a
coeval age of 3 Myr for
all save the BCAH tracks, although it generally spans a slightly older age range
than the primary; on the BCAH tracks it appears to be 5$-$9 Myr.

These comparisons of the derived parameters of the primary and secondary
with the HRDs indicate that effective temperatures better than $\pm 100^\circ$ K
and mass precisions better than $\pm 10 \%$ ~ are required to 
discriminate among the models more definitively than possible at present.
The latter can be accomplished by an improvement of the astrometric
orbit.  This requires only time and patience.   We have started a project
to improve the effective temperature determinations by a comparison
of high resolution spectra of the stars with synthetic spectra following
an approach similar to that of Rice et al. (2010) applied to brown dwarfs.
For main sequence stars of spectral type late G to early M we reach
$T_{eff}$~uncertainties of $\pm 50$K and will explore whether this
accuracy can be achieved for the more active atmospheres of PMS stars.

\section{Summary}

On the basis of our analysis of data taken at the Keck II telescope using the interferometer, adaptive optics imaging,
adaptive optics ``interferometry'' with non-redundant masking, and
high resolution IR spectroscopy of the NTTS 045251$+$3016 binary, we present the
following summary and conclusions:\\

1)  We have determined the NTTS 045251$+$3016 orbital parameters as a resolved 
visual binary and double-lined spectroscopic binary.\\
2)  The masses of the primary and secondary stars are $0.86\pm 0.11$
and $0.55\pm 0.05$ ~\msun, respectively.\\
3)  The distance to the system is $158.7\pm 3.9$ ~pc, placing it
on the far side of the Taurus-Auriga SFR.\\
4)  The measured masses and distance differ significantly from
the values determined by Steffen et al. (2001).   Their orbit is 
in error possibly because their astrometry with the FGS on the HST
relied on measurements from the FGS at the limits of its sensitivity.\\
5) By determining the primary and secondary star absolute H or K
magnitude or luminosity and $T_{\rm eff}$ and plotting their locations
on HRDs, we compare the evolutionary tracks calculated by Baraffe et al. 
(1998), Siess et al. (2000), Tognelli et al. (2011), and the Dartmouth 
group (Feiden et al. 2011).  We find that these tracks are mostly
consistent within the uncertainties compared to the measured dynamical mass
of the primary and indicate an age $\sim 3$ Myr.   The secondary star
dynamical mass is least consistent with the BCAH tracks.   It appears older 
than 3 Myr on all the tracks but, given the uncertainties in $T_{\rm eff}$, 
it would be premature to believe that the components are not coeval. \\
6) Improvement in comparisons of empirical data for stars
of mass less than $\sim1$ \msun~with the 
several theoretical calculations of evolution to the main sequence
will require precisions on the mass of better than $\pm 10 \%$,
and $\pm 100$ K on $T_{\rm eff}$. 
 
\parindent=0.5in

\vskip 1.0cm

MS thanks Anand Sivaramakrishnan for advice concerning analysis of NRM data 
and Peter Tuthill for information about the implementation of NRM at
Keck.  We thank C. Beichman for the opportunity to obtain an additional 
NIRSPEC spectrum on the night of UT 2010 Nov 22.  We are grateful to Rachel 
Akeson helpful discussions on 
analyzing and interpreting the KI data. We thank the staff at Keck
Observatory for their superb support during and after our use of the 
several instruments.  Data presented herein were obtained at the W. M. Keck
Observatory from telescope time allocated to the National Aeronautics and Space
Administration through the agency's scientific partnership with the California 
Institute of Technology and the University of California.
Keck telescope time was also granted by NOAO, through the Telescope System
Instrumentation Program (TSIP). TSIP is funded by NSF. The Observatory was made
possible by the generous financial support of the W. M. Keck
Foundation.  We recognize the Hawaiian community for
the opportunity to conduct these observations from the summit
of Mauna Kea.  The Keck Interferometer was funded by the National
Aeronautics and Space Administration (NASA) as part of its Exoplanet 
Exploration program.  This work has made use of software produced by the 
NASA Exoplanet Science Institute (NExScI) at the California Institute of 
Technology.  This research has also made use of the Keck Observatory Archive (KOA), 
which is operated by the W. M. Keck Observatory and NExScI, under contract 
with the NASA. Our work was supported in part by NSF Grants AST-09-08406 (MS) 
and AST-1009136 (LP).  GHS acknowledges support from NASA Keck PI Data 
Award administered by NExScI (JPL contract 1441975).

\begin{deluxetable}{llll}
\tabletypesize{\scriptsize}
\tablecaption{Log of Observations \label{tbl-1}}
\tablewidth{0pt}
\tablehead{
\colhead{UT Date} & \colhead{Instrument}  &\colhead{Filter} &\colhead{Calibrators/PSF Stars}
   }
\startdata
2002 Oct 24  & Keck Interferometer & K & HD 27741, 27777, 29645 \\
2002 Nov 21  & Keck Interferometer & K & HD 27777  \\
2004 Jan 09  & Keck Interferometer & K & HD 27741, 27777, 35076  \\
2005 Oct 25  & Keck Interferometer & K & HD 27741, 29645   \\
2006 Nov 11  & Keck Interferometer & K & HD 27741, 27282   \\
2006 Dec 08  & Keck Interferometer & K & HD 27741, 27282   \\
2007 Oct 28  & Keck Interferometer & K & HD 27741  \\
             &                     &      &   \\
2008 Dec 17  & NIRC2+AO            & Hcont, Kcont & DN Tau  \\
2009 Oct 25  & NIRC2+AO            & Hcont, Kcont & DN Tau  \\
2011 Jan 24  & NIRC2+AO            & Hcont        & LkCa 19  \\
2011 Oct 12  & NIRC2+AO            & Hcont        & LkCa 19  \\
             &                     &      &  \\
2011 Jan 24  & NIRC2+AO+NRM        & H    & LkCa 19  \\
2011 Oct 12  & NIRC2+AO+NRM        & H    & LkCa 19  \\
             &                     &      &    \\
2001 Jan 05  & NIRSPEC             & $1.55 \mu$m & \\
2001 Dec 31  & NIRSPEC             & $1.55 \mu$m & \\
2002 Feb 05  & NIRSPEC             & $1.55 \mu$m & \\
2004 Dec 26  & NIRSPEC             & $1.55 \mu$m & \\
2006 Dec 07  & NIRSPEC             & $1.55 \mu$m & \\
2006 Dec 14  & NIRSPEC             & $1.55 \mu$m & \\
2009 Dec 06  & NIRSPEC             & $1.55 \mu$m & \\
2010 Nov 22  & NIRSPEC             & $1.55 \mu$m & \\
\enddata
\end{deluxetable}

\clearpage

\begin{deluxetable}{lcccc}
\tabletypesize{\scriptsize}
\tablewidth{0pt}
\tablecaption{Adopted Angular Diameters of KI Calibrators}
\tablehead{\colhead{Calibrator} & \colhead{Spectral} & \colhead{V} & \colhead{K} & \colhead{Diameter}\\
\colhead{Name} & \colhead{Type} & \colhead{(mag)} & \colhead{(mag)} & \colhead{(mas)}}
\startdata
HD 27282 & G8V & 8.5 & 6.8 & $0.199 \pm 0.014$ \\
HD 27741 & G0V & 8.3 & 6.8 & $0.196 \pm 0.014$ \\
HD 27777 & B8V & 5.7 & 5.8 & $0.172 \pm 0.012$ \\
HD 29645 & G0V & 6.0 & 4.6 & $0.533 \pm 0.037$ \\
HD 35076 & B9V & 6.4 & 6.5 & $0.127 \pm 0.009$ \\
\enddata
\tablecomments{Spectral types and V magnitudes from Kharchenko \& Roeser (2009), K magnitudes from 2MASS (Cutri et al. 2003), angular diameters from SearchCal (Bonneau et al. 2006, 2011).}
\end{deluxetable}

\clearpage

\begin{deluxetable}{ccccccccc}
\tabletypesize{\scriptsize}
\tablewidth{0pt}
\tablecaption{KI $V^2$ Measurements}
\tablehead{\colhead{} & \colhead{UT} & \colhead{} & \colhead{$u$} & \colhead{$v$} & \colhead{$\lambda$} & \colhead{} & \colhead{}  & \colhead{} \\
\colhead{MJD} & \colhead{Date} & \colhead{UTC} & \colhead{(m)} & \colhead{(m)} & \colhead{($\mu$m)} & \colhead{$V^2$} & \colhead{$\sigma_{V^2}$} & \colhead{Fit}}
\startdata
52571.52737 &  2002 Oct 24 &  12:39:24 & 54.086 & 63.082 & 2.18 & 0.658 & 0.037 &   \\
52571.52956 &  2002 Oct 24 &  12:42:33 & 53.854 & 63.458 & 2.18 & 0.605 & 0.043 &   \\
52599.52655 &  2002 Nov 21 &  12:38:13 & 40.508 & 74.691 & 2.18 & 0.712 & 0.063 &   \\
52599.52824 &  2002 Nov 21 &  12:40:39 & 40.086 & 74.908 & 2.18 & 0.758 & 0.062 &   \\
53013.38074 &  2004 Jan 09 &  09:08:15 & 43.537 & 72.987 & 2.18 & 0.655 & 0.089 &   \\
53668.56213 &  2005 Oct 25 &  13:29:27 & 48.617 & 69.341 & 2.15 & 0.567 & 0.036 & * \\
53668.56353 &  2005 Oct 25 &  13:31:28 & 48.361 & 69.557 & 2.15 & 0.560 & 0.026 & * \\
53668.59009 &  2005 Oct 25 &  14:09:43 & 42.806 & 73.422 & 2.15 & 0.400 & 0.019 & * \\
53668.59141 &  2005 Oct 25 &  14:11:37 & 42.497 & 73.601 & 2.15 & 0.384 & 0.021 & * \\
53668.60165 &  2005 Oct 25 &  14:26:22 & 40.004 & 74.948 & 2.15 & 0.335 & 0.023 & * \\
53668.60296 &  2005 Oct 25 &  14:28:15 & 39.674 & 75.113 & 2.15 & 0.329 & 0.024 & * \\
53668.61347 &  2005 Oct 25 &  14:43:23 & 36.922 & 76.395 & 2.15 & 0.257 & 0.019 & * \\
53668.61526 &  2005 Oct 25 &  14:45:58 & 36.437 & 76.604 & 2.15 & 0.249 & 0.020 & * \\
53668.61974 &  2005 Oct 25 &  14:52:25 & 35.200 & 77.116 & 2.15 & 0.255 & 0.021 & * \\
53668.62150 &  2005 Oct 25 &  14:54:57 & 34.708 & 77.312 & 2.15 & 0.250 & 0.021 & * \\
54050.56118 &  2006 Nov 11 &  13:28:06 & 38.664 & 75.604 & 2.18 & 0.537 & 0.044 & * \\
54077.46703 &  2006 Dec 08 &  11:12:31 & 43.644 & 72.922 & 2.18 & 0.521 & 0.041 &   \\
54077.51305 &  2006 Dec 08 &  12:18:47 & 31.546 & 78.471 & 2.18 & 0.786 & 0.087 &   \\
54401.63071 &  2007 Oct 28 &  15:08:13 & 30.045 & 78.963 & 2.18 & 0.605 & 0.107 & * \\
54401.63225 &  2007 Oct 28 &  15:10:26 & 29.579 & 79.109 & 2.18 & 0.630 & 0.113 & * \\
\enddata
\tablecomments{An asterisk in the last column indicates that the measurement was included in the orbit fit.}
\label{tab.v2}
\end{deluxetable}

\begin{deluxetable}{lllllcccc}
\tabletypesize{\scriptsize}
\tablecaption{Log of NIRC2+AO Observations\label{tbl-3}}
\tablewidth{0pt}
\tablehead{
\colhead{UT Date} & \colhead{Target} &\colhead{UTC} &\colhead{Mode} &\colhead{Filter} & \colhead{AO Rate} & \colhead{$T_{\rm int}$(s)} & \colhead{Co-adds}&\colhead{Images}
}
\startdata
2008 Dec 17  & NTTS 045251$+$3016 & 09:03  & AO & Kcont  & 250 &  1.5 & 10 & 12  \\
             & NTTS 045251$+$3016 & 09:11  & AO & Hcont  & 250 &  1.2 & 10 & 12  \\
             & DN Tau           & 09:24  & AO & Hcont  & 250 &  0.3 & 10 &  6  \\
             & DN Tau           & 09:26  & AO & Hcont  & 250 &  1.0 & 10 &  6  \\
             & DN Tau           & 09:31  & AO & Kcont  & 250 &  1.0 & 10 & 12  \\
2009 Oct 25  & NTTS 045251$+$3016 & 15:36  & AO & Kcont  & 438 &  0.5 & 10 &  6  \\
             & NTTS 045251$+$3016 & 15:40  & AO & Hcont  & 438 &  0.5 & 10 &  6  \\
             & DN Tau           & 15:44  & AO & Hcont  & 438 &  0.2 & 10 &  3  \\
             & DN Tau           & 15:47  & AO & Kcont  & 438 &  0.2 & 10 &  3  \\
2011 Jan 24  & LkCa 19          & 08:57  & AO & Hcont  & 438 &  0.3 & 10 &  6  \\
             & NTTS 045251$+$3016 & 09:01  & AO & Hcont  & 438 &  0.3 & 10 & 12  \\
             & NTTS 045251$+$3016 & 09:20  & AO+NRM & H  & 438 & 20.0 &  1 & 16  \\
             & LkCa 19          & 09:35  & AO+NRM & H  & 438 & 20.0 &  1 & 16  \\
             & NTTS 045251$+$3016 & 09:53  & AO+NRM & H  & 438 & 15.0 &  1 & 16  \\
             & LkCa 19          & 10:05  & AO+NRM & H  & 438 & 15.0 &  1 & 16  \\
2011 Oct 12  & NTTS 045251$+$3016 & 14:12  & AO & Hcont  & 438 &  0.5 & 10 & 12 \\
             & NTTS 045251$+$3016 & 14:18  & AO & Kcont  & 438 &  0.5 & 10 & 12 \\
             & LkCa 19          & 14:23  & AO & Hcont  & 438 &  0.5 & 10 &  6 \\
             & LkCa 19          & 14:28  & AO & Kcont  & 438 &  0.5 & 10 &  6 \\
             & NTTS 045251$+$3016 & 14:48  & AO+NRM & H  & 438 &  5.0 &  1 & 16 \\
             & LkCa19           & 14:53  & AO+NRM & H  & 438 &  5.0 &  1 &  8 \\
             & LkCa19           & 14:57  & AO+NRM & H  & 438 &  4.0 &  1 &  8 \\
             & NTTS 045251$+$3016 & 15:05  & AO+NRM & H  & 438 &  4.0 &  1 & 16 \\
             & LkCa19           & 15:11  & AO+NRM & H  & 438 &  4.0 &  1 & 16 \\
\enddata
\end{deluxetable}

\clearpage

\begin{deluxetable}{llllll}
\tabletypesize{\scriptsize}
\tablecaption{Astrometric Results \label{tbl-4}}
\tablewidth{0pt}
\tablehead{
\colhead{UT Date} & \colhead{MJD} &\colhead{Instrument} &\colhead{$\rho$(mas)} &\colhead{PA(deg)} &\colhead{Flux Ratio}}
\startdata
2008 Dec 17  & 54817.383 & NIRC2+AO      & $33.0 \pm 1.3$ & $175.4 \pm 2.3$  & $0.385 \pm 0.032$ (K), $0.418 \pm 0.037$ (H)   \\
2009 Oct 25  & 55129.653 & NIRC2+AO      & $35.9 \pm 2.5$ & $179.8 \pm 4.1$  & $0.408 \pm 0.036$ (K), $0.488 \pm 0.037$ (H)   \\
2011 Jan 24  & 55585.375 & NIRC2+AO      & $27.6 \pm 3.8$ & $182.4 \pm 7.9$  & $0.445 \pm 0.103$ (H)  \\
             &           &               &                &                  &     \\
2008 Dec 23\tablenotemark{a} & 54823.528   & NIRC2+AO+NRM  & $32.2 \pm 0.1$ & $180.8 \pm 1.8$  & $0.41 \pm 0.01$ (K)    \\
2011 Jan 24  & 55585.412 & NIRC2+AO+NRM  & $27.4 \pm 0.3$ & $186.2 \pm 2.7$  & $0.44 \pm 0.01$ (H)   \\
2011 Oct 12  & 55846.617 & NIRC2+AO+NRM  & $16.2 \pm 1.4$ & $194.9 \pm 2.5$  & $0.27 \pm 0.02$ (H)   \\
\enddata
\tablenotetext{a}{Analysis of NRM data published by Kraus et al. (2011); see text for discussion.}
\end{deluxetable}

\begin{deluxetable}{ccrrrrc}
\tabletypesize{\scriptsize}
\tablecaption{Radial Velocities of the Primary and Secondary \label{tbl-5}}
\tablewidth{0pt}
\tablehead{
\colhead{}&\colhead{UT}&\colhead{V$_1$}&\colhead{$\sigma_1$}&\colhead{V$_2$}&\colhead{$\sigma_2$}&\colhead{}\\
\colhead{MJD}&\colhead{Date}&\colhead{km s$^{-1}$}&\colhead{km s$^{-1}$}&\colhead{km s$^{-1}$}&\colhead{km s$^{-1}$}&\colhead{Phase}}
\startdata
51914.311& 2001.012& 14.68 & 1.00 & 14.80 &  2.00  &   0.102\\
52274.344& 2001.998& 18.85 & 1.00 & 7.34   &  2.00  &  0.244 \\
52310.313& 2002.097& 18.90 & 1.00 & 6.50   &  2.00  & 0.258 \\
53365.428& 2004.985& 16.14 & 1.00 & 13.50 & 2.00 &  0.674\\
54076.537& 2006.933& 4.39  &  1.00 & 31.69 & 2.00 & 0.954\\
54083.025& 2006.953&   2.70 & 1.00 & 29.42 &  2.0  & 0.957\\
55171.395& 2009.930& 18.48 & 1.00 & 7.60   & 2.00 & 0.385\\ 
55522.592& 2010.892& 17.64 & 1.00 & 11.65 & 2.00 & 0.523\\ 
\enddata
\end{deluxetable}

\clearpage
\begin{deluxetable}{ll}
\tabletypesize{\scriptsize}
\tablecaption{Orbital Parameters, Masses, and Distance \label{tbl-6}}
\tablewidth{0pt}
\tablehead{
\colhead{Parameter}&\colhead{Value}
}
\startdata
$P$ (days)           & $2513.8  \pm 2.9$       \\
$T$  (MJD)           & $51626.3 \pm 9.9$       \\
$e$                  & $0.4937  \pm 0.0071$    \\
$a$ (mas)            & $25.57   \pm 0.12$      \\
$i$ ($^\circ$)        & $78.1    \pm 1.0$        \\
$\Omega$ ($^\circ$)   & $178.06  \pm 0.77$        \\
$\omega$ ($^\circ$)   & $213.0   \pm 1.7$         \\
$K_1$    (km/s)      & $7.67    \pm 0.14$       \\ 
$K_2$    (km/s)      & $12.09   \pm 0.44$       \\ 
$\gamma$ (km/s)      & $14.43   \pm 0.09$      \\ 
KI flux ratio at $K$ & $ 0.389   \pm 0.042$     \\
                     &     \\
$M_1$ (\msun)        & $0.86 \pm 0.11$  \\           
$M_2$ (\msun)        & $0.55 \pm 0.05$     \\
$q = M_2/M_1$        & $0.635 \pm 0.026$ \\
$d$   (pc)           & $158.7 \pm 3.9$     \\
                     &      \\
$\chi^2$             &   98.7   \\
$\chi^2_\nu $         &   1.05   \\
\enddata
\end{deluxetable}

\begin{deluxetable}{lcccc}
\tabletypesize{\scriptsize}
\tablewidth{0pt}
\tablecaption{Comparison of Dynamical and Track Masses}
\tablehead{\colhead{Source} & \colhead{$M_1$ (M$_{\odot}$)} & \colhead{Age (Myr)} & \colhead{$M_2$ (M$_{\odot}$)} & \colhead{Age (Myr)}}
\startdata
This Work              & 0.86$\pm$0.11        & \nodata & 0.55$\pm$0.05        & \nodata  \\
Baraffe et al. (1998)  & 1.05$^{+0.20}_{-0.10}$ & 3$\pm$1 & 0.70$^{+0.15}_{-0.05}$ & 8$\pm$3  \\
Siess et al. (2000)    & 1.10$^{+0.15}_{-0.20}$ & 3$\pm$2 & 0.60$^{+0.20}_{-0.10}$ & 4$\pm$3 \\
Tognelli et al. (2011) & 0.95$\pm$0.15\tablenotemark{a} & 3$\pm$2\tablenotemark{a} & 0.70$^{+0.20}_{-0.10}$\tablenotemark{b} & 5$\pm$3\tablenotemark{b}  \\
Feiden et al. (2011)   & 1.00$\pm$0.15        & 3$\pm$1 & 0.65$^{+0.20}_{-0.10}$ & 5$\pm$3  \\
\enddata
\tablenotetext{a}{Mixing length parameter $=$ 1.68.}
\tablenotetext{b}{Mixing length parameter $=$ 1.20.}
\end{deluxetable}
\clearpage

\begin{figure}
\begin{center}
\includegraphics[scale=0.67]{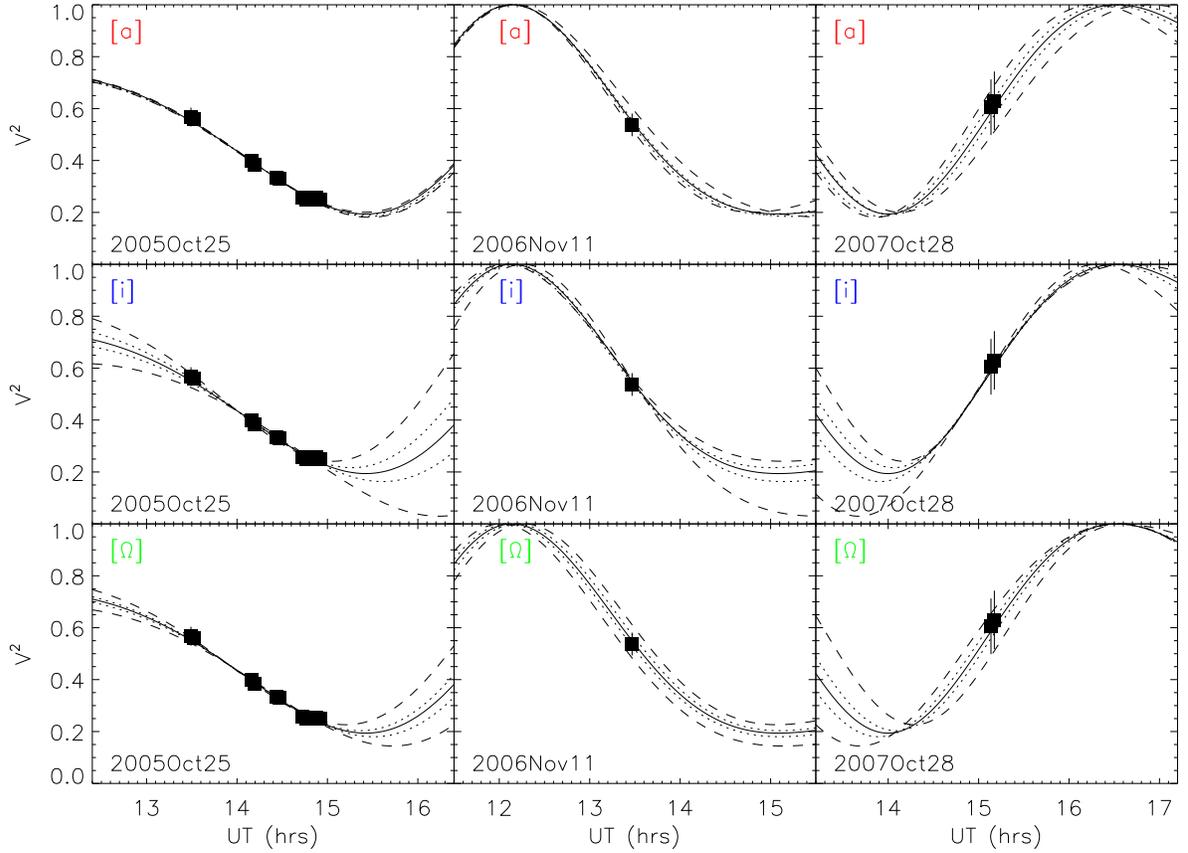}
\end{center}
\caption{
Calibrated visibilities measured with the Keck Interferometer on UT
2005 October 25, 2006 November 11, and 2007 October 28.  The solid lines 
shows the simultaneous orbit fit to the RV, AO, NRM, and KI data.  In each 
panel, we show how the fit changes if we vary the semi-major axis (top row), 
inclination (middle row), and $\Omega$ (bottom row) by their 1\,$\sigma$ 
(dotted line) and 3\,$\sigma$ (dashed line) uncertainties while optimizing
the remaining parameters.  The parameter
being varied is identified in brackets in the top left corner of each plot.  
The dashed orbits that represent the 3\,$\sigma$ intervals for $a, i,$ and
$\Omega$ are also plotted in Figure 5 to show how they compare with the AO
and NRM measurements.
}
\end{figure}
\clearpage

\begin{figure}
\begin{center}
\includegraphics[scale=0.9]{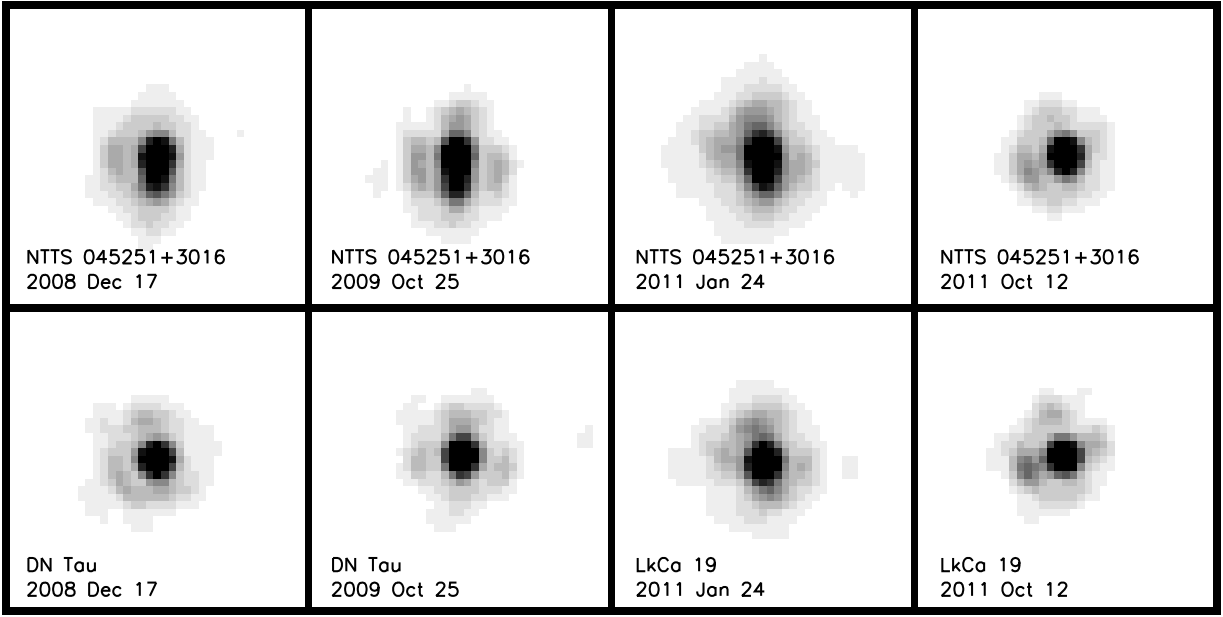}
\end{center}
\caption{
H-band images of NTTS 045251$+$3016 (top) and the point-spread function 
(PSF) calibrators DN Tau and LkCa 19 (bottom) obtained with NIRC2
operating with Adaptive Optics on UT 2008 Dec 17, 2009 Oct 25, 2011 Jan
24, and 2011 Oct 12.  The AO images of NTTS 0455251$+$3016 in 2008,
2009, and Jan 2011 resolved the binary  but those taken in Oct 2011 
when NRM observations revealed that the separation was $\sim
16$ mas (Table 4) did not reliably resolve the pair. Each panel is 400 mas wide.
}
\end{figure}
\clearpage

\begin{figure}
\begin{center}
\includegraphics[scale=0.6]{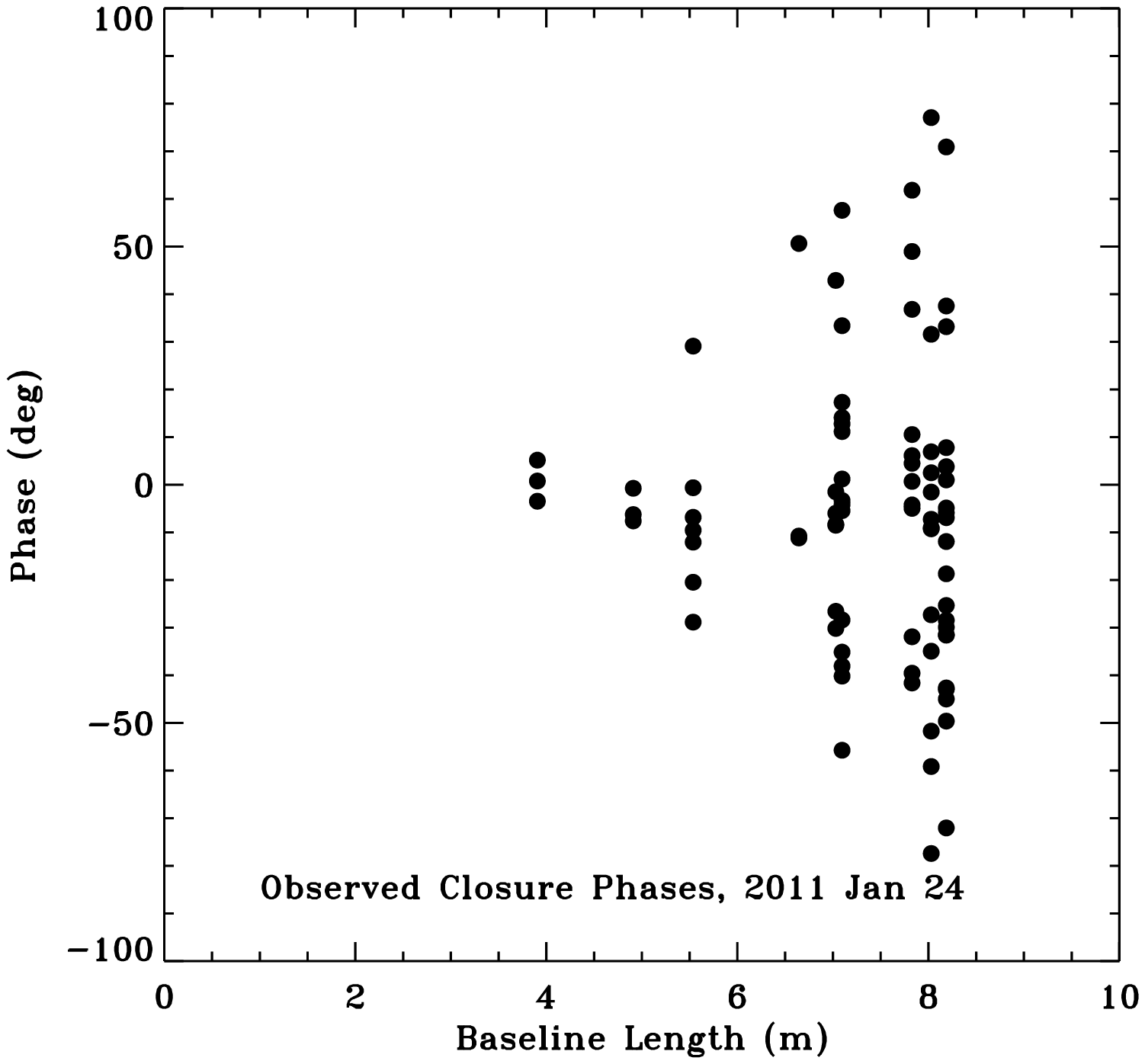} \\
\includegraphics[scale=0.6]{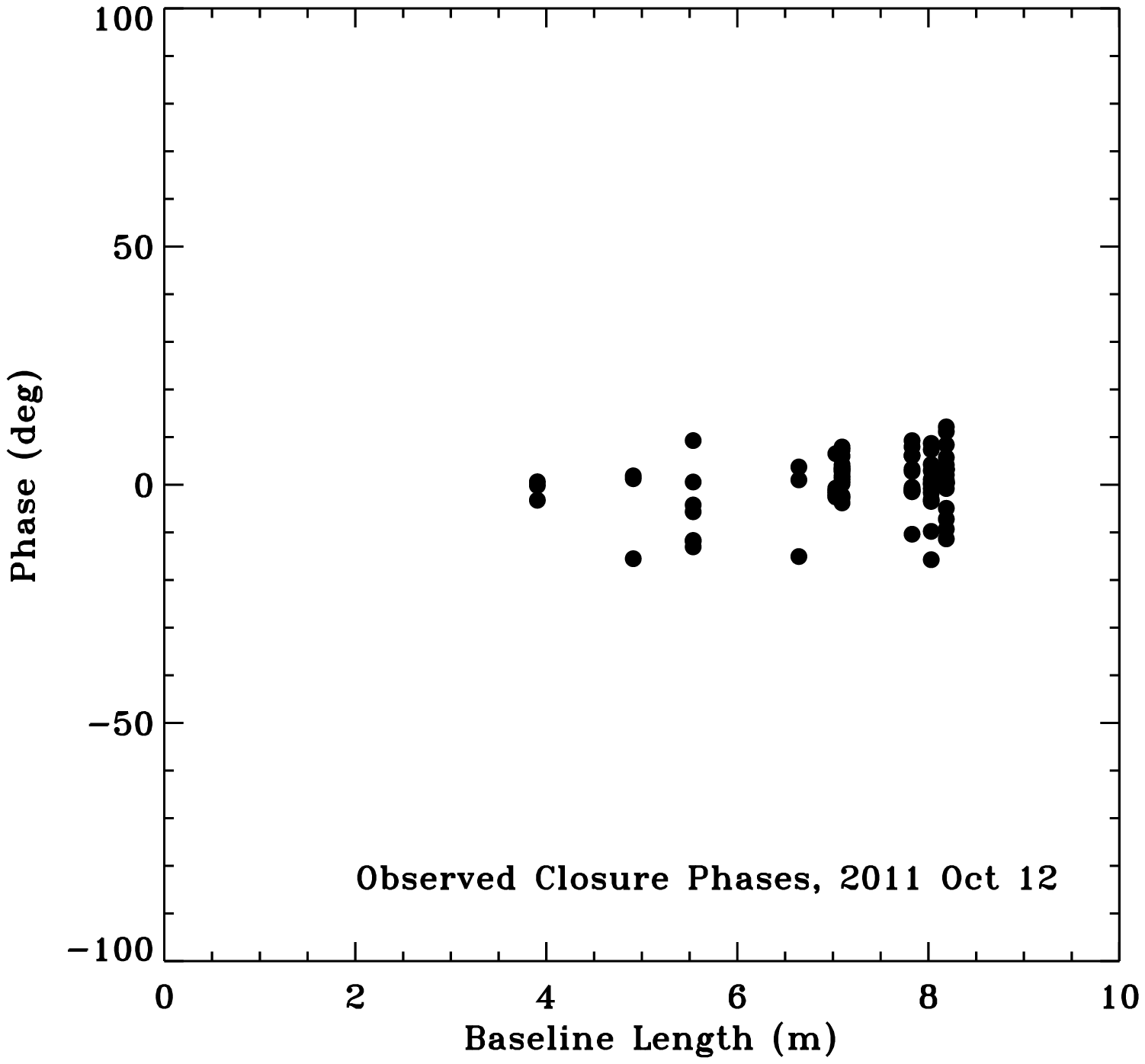}
\end{center}
\caption{Closure phase (degrees) {\it vs} longest baseline (meters) in the triangle
for which it is measured for NRM data obtained on UT 2011 Jan 24 ({\it top}) and 
2011 Oct 12 ({\it below}).  For each data set, average closure phases
are shown only for the first quarter of each data set (see text).
The binary separation, position angle, and flux ratio of the models
that fit each data set best are reported in Table 4.  
}
\end{figure}
\clearpage

\begin{figure}
\begin{center}
\includegraphics[scale=0.8]{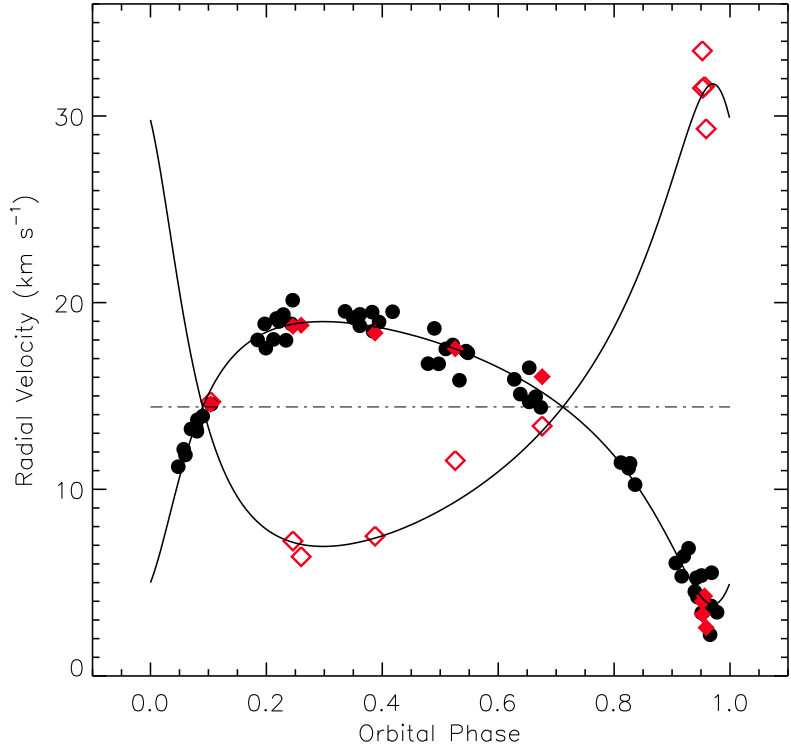}
\end{center}
\caption{
RV as a function of phase for the primary (filled
symbols) and secondary (open symbols) stars in NTTS 04251$+$3016.  
The black circles represent Steffen et al.'s (2001) measurements of the 
primary RV in visible light.  The red symbols were measured in
IR light. K5V and M0V spectral templates were used to measure
the RVs derived from the IR spectra (\S 2.4).   
The two IR measurements at phase 0.953 and 9.954
were reported previously by Steffen et al. 
}
\end{figure}
\clearpage

\begin{figure}
\begin{center}
\includegraphics[scale=0.65]{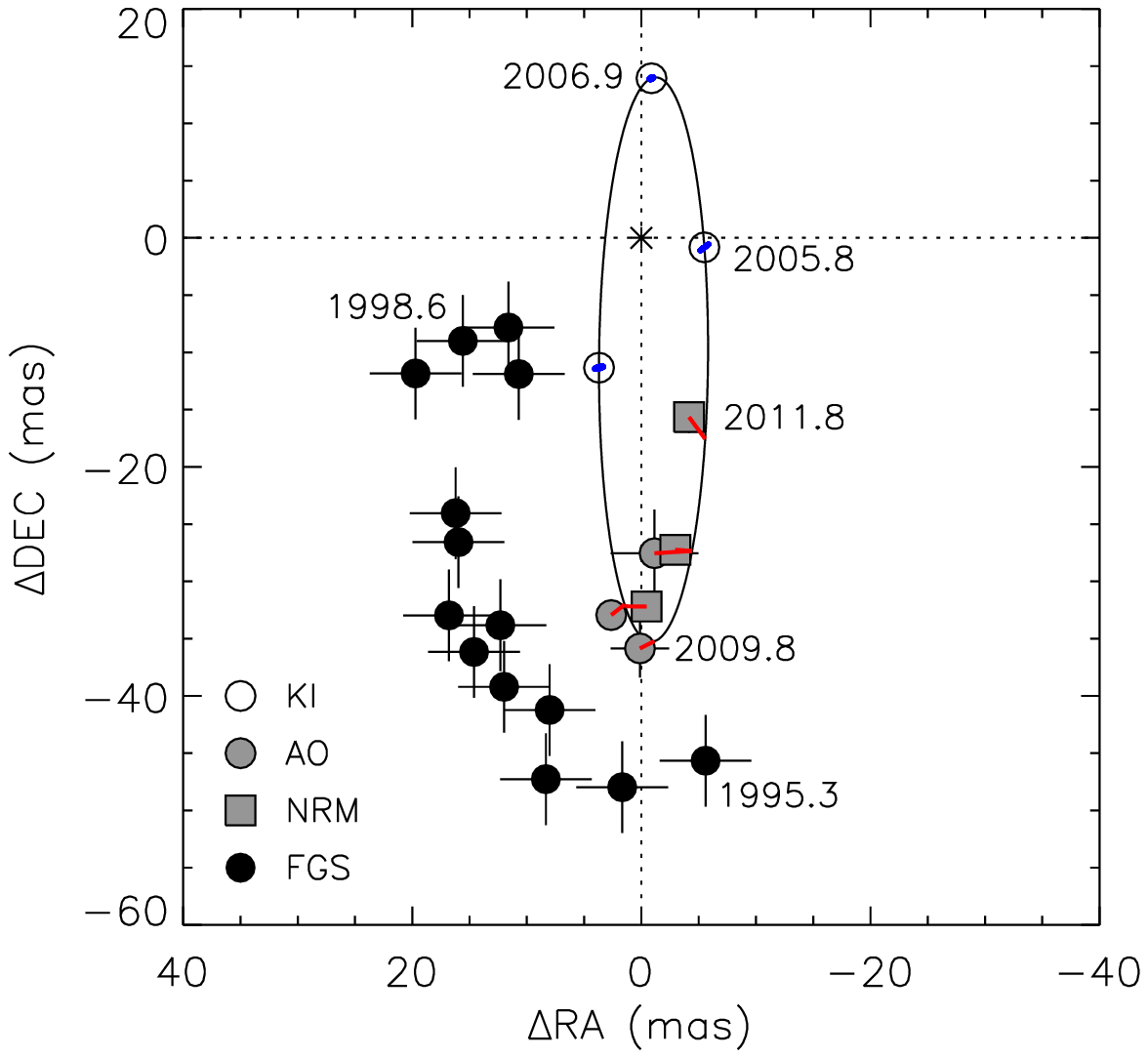}
\includegraphics[scale=0.65]{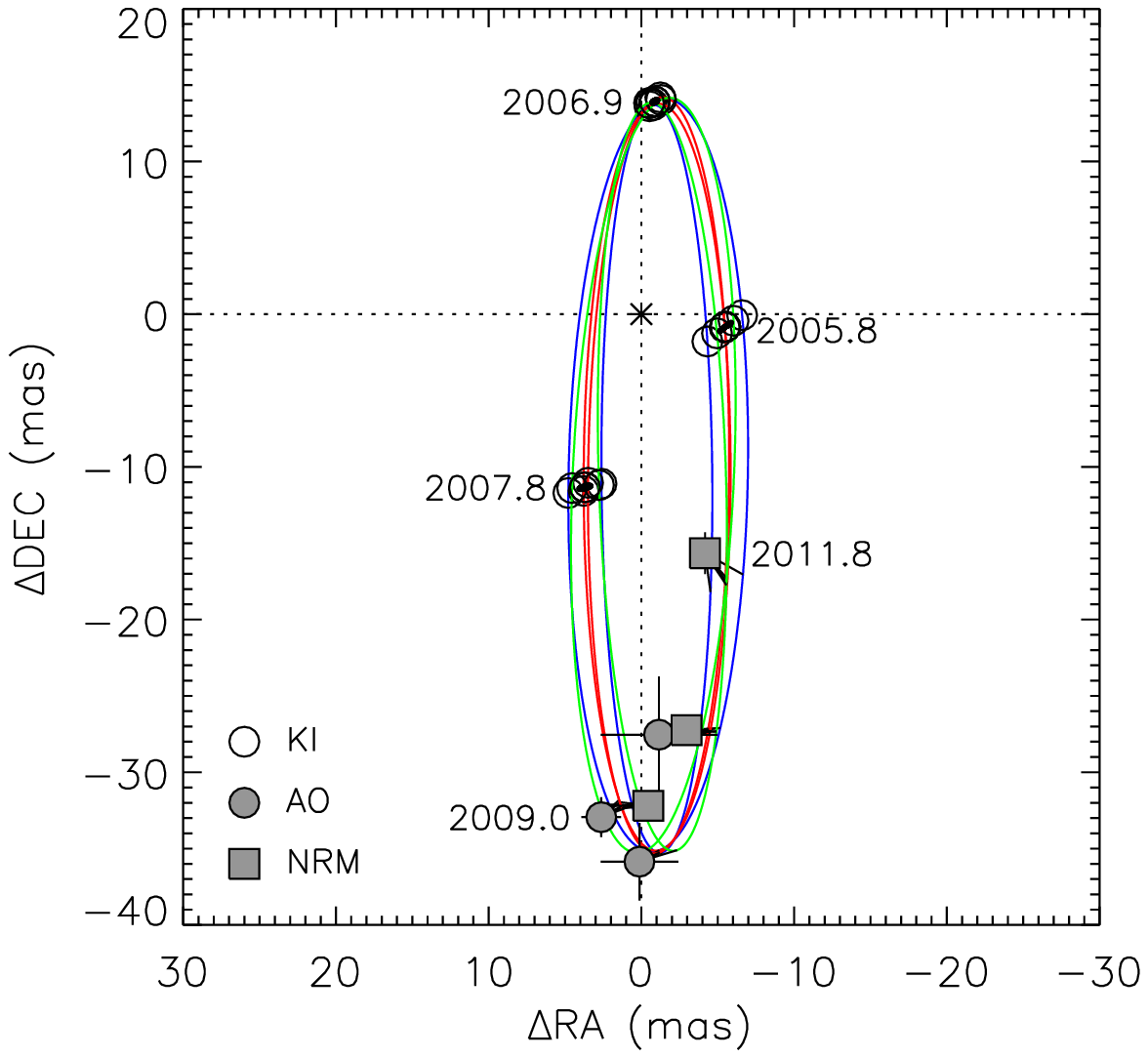}
\end{center}
\caption{
{\it Left}: Visual orbit of NTTS 04251$+$3016 based on the fit to 
the AO, NRM, KI, and RV measurements.  The shaded grey circles indicate the 
AO data and the shaded grey squares represent the NRM measurements.  The red 
lines connect the measured values with the computed positions from the orbit 
fit.  The computed positions of the binary during the epochs 
of the KI observations are marked by the open circles.  The small blue 
ellipses within the open circles show the variation in $\Delta$RA and 
$\Delta$DEC during the times of the KI epochs that are generated by exploring 
the range of orbital solutions that fit the data within the 1\,$\sigma$ 
($\Delta \chi^2 = 1$) confidence interval.
The figure also includes the FGS measurements (black circles) 
reported by Steffen et al. (2001).   They are obviously
inconsistent with the new observations reported here.  We discuss the
possible reason for the discrepancy in \S 3.1.
{\it Right}: Range of orbits obtained by varying $a$ (red), $i$ (blue), and 
$\Omega$ (green) by their 3\,$\sigma$ uncertainties while optimizing the 
remaining parameters.  These are the same orbits that are plotted as dashed 
lines in Figure 1 and illustrate how varying the orbital parameters 
affects simultaneously the fits to the KI visibilities and the AO/NRM 
measurements.  The computed position of the binary during the epochs of the 
KI observations (open circles) varies depending on the orbital parameters.
}
\end{figure}
\clearpage

\begin{figure}
\begin{center}
\includegraphics[scale=0.6]{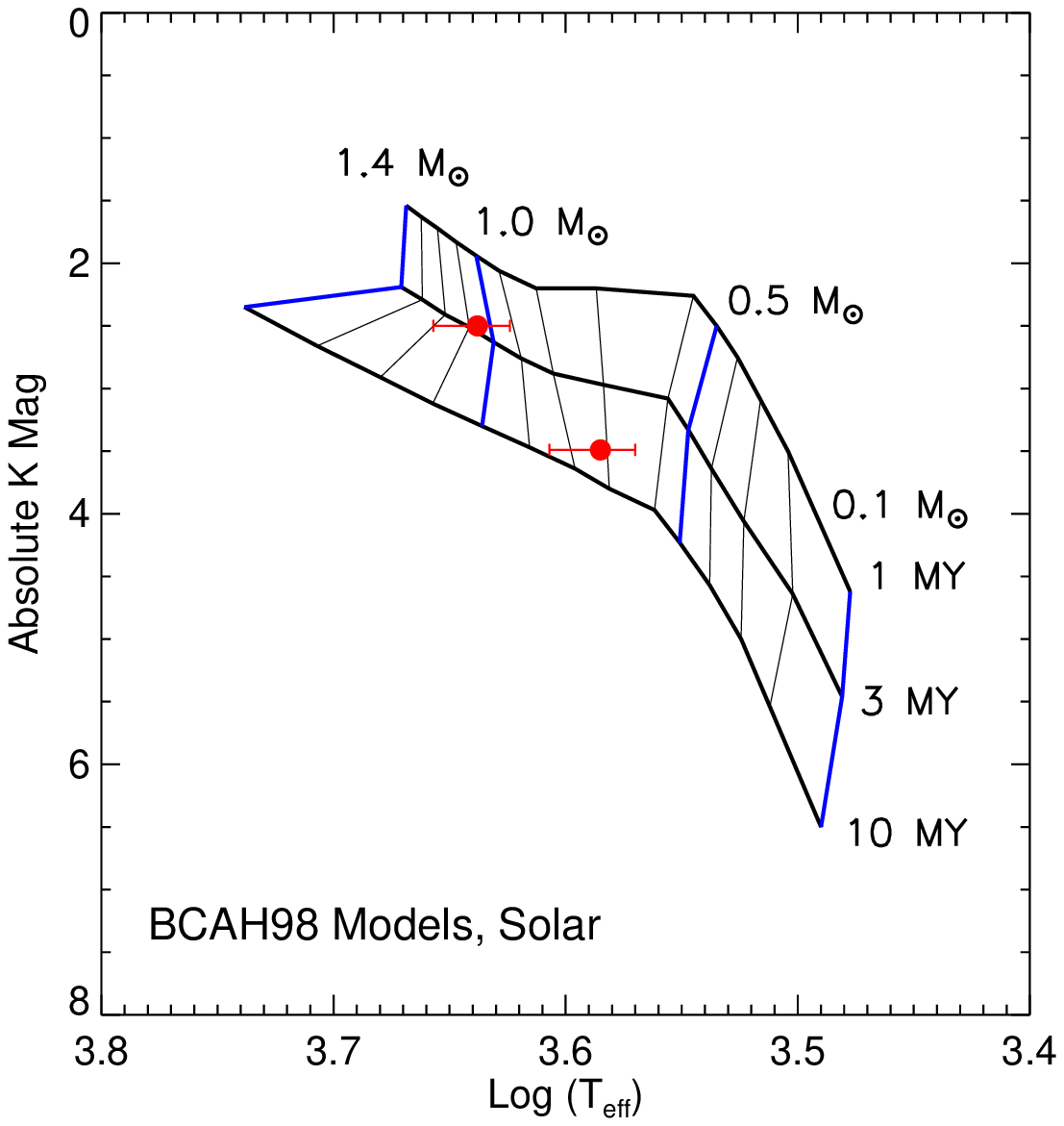}
\includegraphics[scale=0.6]{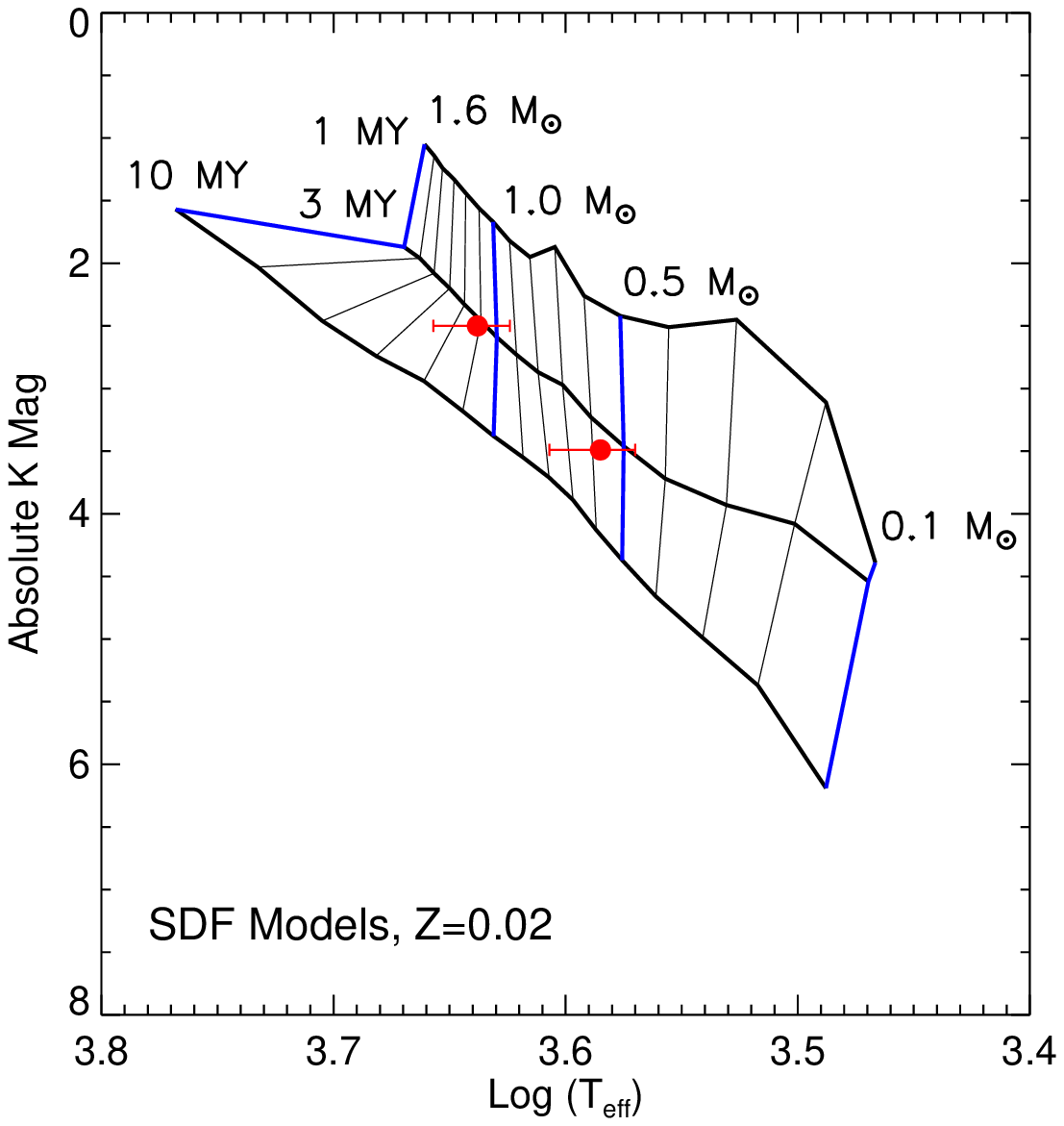}
\includegraphics[scale=0.6]{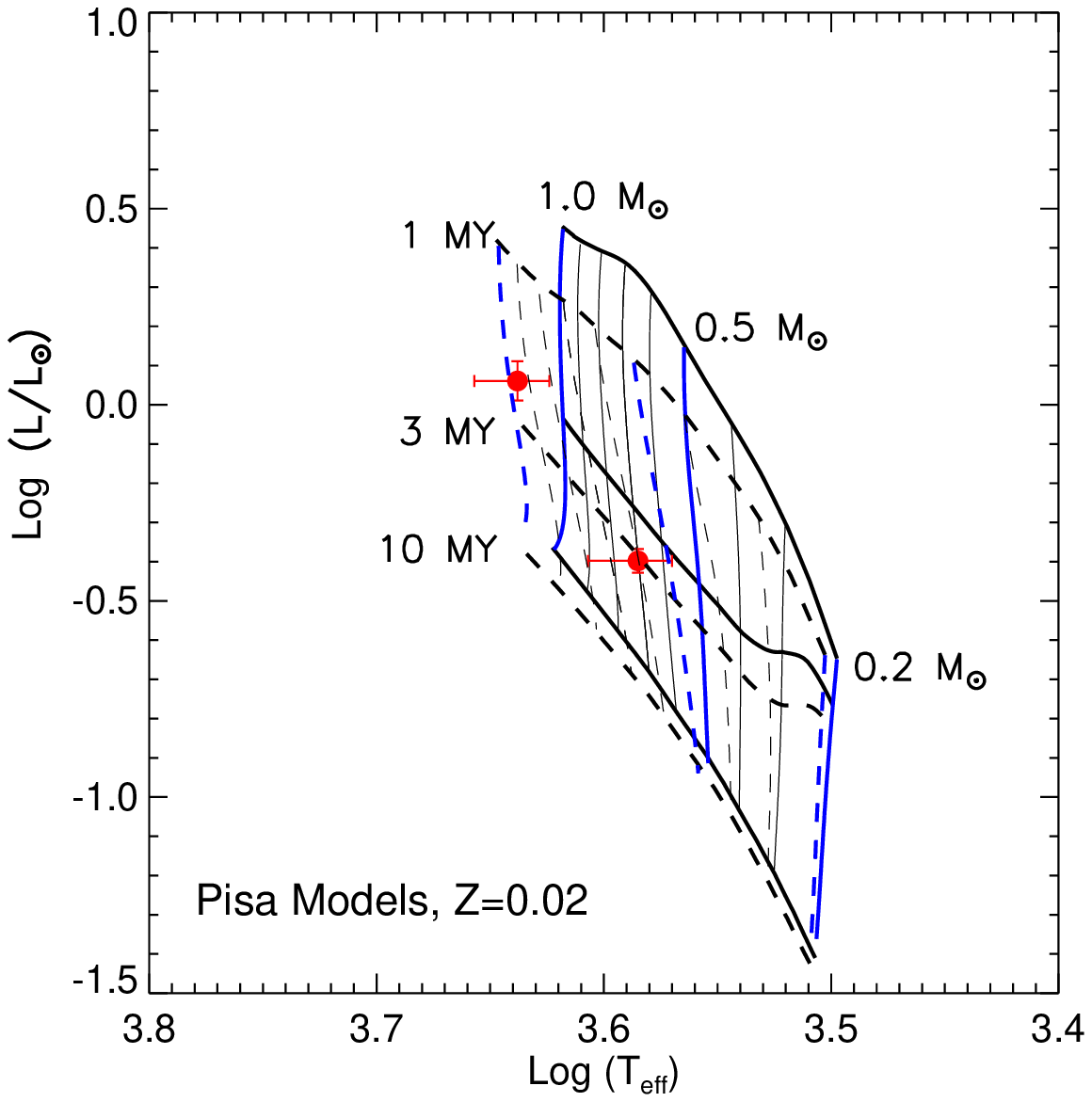}
\includegraphics[scale=0.6]{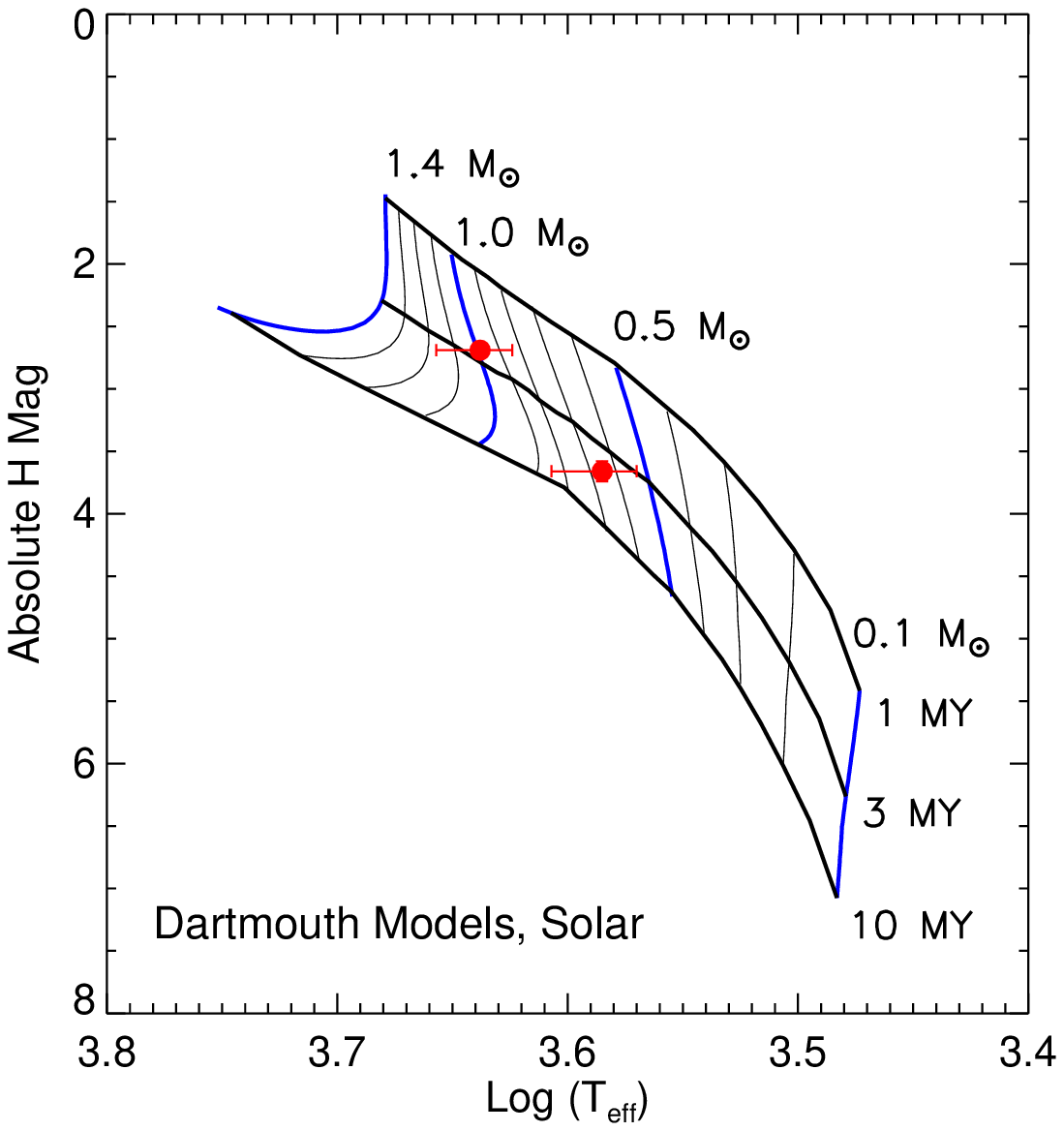}

\end{center}
\caption{\footnotesize HRDs showing the location of the primary and secondary in
NTTS 045251$+$3016 with respect to theoretical PMS evolutionary
tracks and isochrones.
The vertical uncertainties include those in photometry, flux ratio,
and distance (see \S 3.3).
{\it Top left}: Evolutionary tracks of BCAH at solar
metallicity for masses of 0.1 to 1.4 \msun. For masses below 0.6
\msun, we used the tracks calculated with a mixing length parameter 
$\alpha =1.0$; for larger masses we used the tracks with $\alpha=1.9$.    
This HRD uses the absolute K-band magnitude, $M_K$, as a proxy for
the luminosity. {\it Top right}: Evolutionary tracks of SDF at 
Z = 0.02 for masses of 0.1 to 1.6 \msun ~again using $M_K$. 
{\it Bottom left}: Pisa stellar evolutionary models (Tognelli et al. 2011) for
$\alpha = 1.20$ (solid lines) and 1.68 (dashed lines) 
at a metallicity of Z = 0.02. 
{\it Bottom right}: The Dartmouth tracks at solar abundances and $\alpha=1.83$,
using $M_H$ (Feiden et al. 2011).
The isochrones are plotted for 1, 3, and 10 Myr.  Labelled mass tracks are plotted
in blue.  Uncertainties in the y-axes are on the order of or smaller than the plot symbols.}
\end{figure}


\begin{thebibliography}

\bibitem[Baraffe et al.(1998)]{1998A&A...337..403B} Baraffe, I., Chabrier, G.,
Allard, F., \& Hauschildt, P.~H.\ 1998, \aap, 337, 403 (BCAH)

\bibitem[Baldwin et al.(1986)]{1986Natur.320..595B} Baldwin, J.~E., Haniff, 
C.~A., Mackay, C.~D., \& Warner, P.~J.\ 1986, \nat, 320, 595

\bibitem[Berger \& Segransan(2007)]{2007NewAR..51..576B} Berger, J.~P., \&
Segransan, D.\ 2007, \nar, 51, 576

\bibitem[Bertout \& Genova(2006)]{2006A&A...460..499B} Bertout, C., \&
Genova, F.\ 2006, \aap, 460, 499

\bibitem[Boden(2000)]{2000plbs.conf....9B} Boden, A.~F. 2000, in Principles of Long Baseline Stellar Interferometry, ed. P.~R. Lawson (Pasadena, CA: JPL), 9, http://olbin.jpl.nasa.gov/iss1999/coursenotes.html

\bibitem[Bonneau et al.(2006)]{2006A&A...456..789B} Bonneau, D., Clausse, J.-M.,
Delfosse, X., et al.\ 2006, \aap, 456, 789

\bibitem[Bonneau et al.(2011)]{2011A&A...535A..53B} Bonneau, D., Delfosse, X.,
Mourard, D., et al.\ 2011, \aap, 535, A53

\bibitem[Bridle \& Schwab(1999)]{1999ASPC..180..371B} Bridle, A.~H., \&
Schwab, F.~R.\ 1999, Synthesis Imaging in Radio Astronomy II, 180, 371

\bibitem[Colavita et al.(2004)]{2004SPIE.5491..454C} Colavita, M.~M., 
Wizinowich, P.~L., \& Akeson, R.~L.\ 2004, \procspie, 5491, 454

\bibitem[Cutri et al.(2003)]{2003yCat.2246....0C} Cutri, R.~M., Skrutskie, 
M.~F., van Dyk, S., et al.\ 2003, VizieR Online Data Catalog, 2246, 0

\bibitem[Dotter et al.(2008)]{2008ApJS..178...89D} Dotter, A., Chaboyer, 
B., Jevremovi{\'c}, D., et al.\ 2008, \apjs, 178, 89

\bibitem[Feiden et al.(2011)]{2011ApJ...740L..25F} Feiden, G.~A., Chaboyer, 
B., \& Dotter, A.\ 2011, \apjl, 740, L25

\bibitem[Herbig \& Bell(1988)]{1988cels.book.....H} Herbig, G.~H., \& Bell, K.~R.\ 1988,
Third catalog of emission-line stars of the Orion population., by G.H.~gerbig and K.R.~Bell.~ Lick
Observatory Bulletin \#1111, Santa Cruz: Lick Observatory, June 1988, 90

\bibitem[Hillenbrand \& White(2004)]{2004ApJ...604..741H} Hillenbrand, L.~A., \&
White, R.~J.\ 2004, \apj, 604, 741

\bibitem[Kenyon et al.(1994)]{1994AJ....108.1872K} Kenyon, S.~J., 
Dobrzycka, D., \& Hartmann, L.\ 1994, \aj, 108, 1872

\bibitem[Kenyon \& Hartmann(1995)]{1995ApJS..101..117K} Kenyon, S.~J., \&
Hartmann, L.\ 1995, \apjs, 101, 117

\bibitem[Kharchenko \& Roeser(2009)]{2009yCat.1280....0K} Kharchenko, N.~V., \&
Roeser, S.\ 2009, VizieR Online Data Catalog, 1280, 0

\bibitem[Kraus et al.(2011)]{2011ApJ...731....8K} Kraus, A.~L., Ireland, 
M.~J., Martinache, F., \& Hillenbrand, L.~A.\ 2011, \apj, 731, 8

\bibitem[Lachaume \& Berger(2012)]{2012SPIE.8445E..3LL} Lachaume, R., \&
Berger, J.-P.\ 2012, \procspie, 8445, 3

\bibitem[Leinert et al.(1993)]{1993A&A...278..129L} Leinert, C., Zinnecker, H.,
Weitzel, N., et al.\ 1993, \aap, 278, 129

\bibitem[Lloyd et al.(2006)]{2006ApJ...650L.131L} Lloyd, J.~P., Martinache, 
F., Ireland, M.~J., et al.\ 2006, \apjl, 650, L131

\bibitem[Luhman et al.(2003)]{2003ApJ...590..348L} Luhman, K.~L., 
Brice{\~n}o, C., Stauffer, J.~R., et al.\ 2003, \apj, 590, 348

\bibitem[Mace et al.(2012)]{2012AJ....144...55M} Mace, G.~N., Prato, L., 
Torres, G., et al.\ 2012, \aj, 144, 55

\bibitem[Martinache et al.(2009)]{2009ApJ...695.1183M} Martinache, F., 
Rojas-Ayala, B., Ireland, M.~J., Lloyd, J.~P., 
\& Tuthill, P.~G.\ 2009, \apj, 695, 1183

\bibitem[McLean et al.(2000)]{2000SPIE.4008.1048M} McLean, I.~S., Graham, 
J.~R., Becklin, E.~E., et al.\ 2000, \procspie, 4008, 1048

\bibitem[Nakajima et al.(1989)]{1989AJ.....97.1510N} Nakajima, T., 
Kulkarni, S.~R., Gorham, P.~W., et al.\ 1989, \aj, 97, 1510

\bibitem[Rice et al.(2010)]{2010ApJS..186...63R} Rice, E.~L., Barman, T., 
Mclean, I.~S., Prato, L., \& Kirkpatrick, J.~D.\ 2010, \apjs, 186, 63

\bibitem[Schaefer et al.(2008)]{2008AJ....135.1659S} Schaefer, G.~H., 
Simon, M., Prato, L., \& Barman, T.\ 2008, \aj, 135, 1659

\bibitem[Schaefer et al.(2012)]{2012ApJ...756..120S} Schaefer, G.~H., 
Prato, L., Simon, M., \& Zavala, R.~T.\ 2012, \apj, 756, 120

\bibitem[Siess et al.(2000)]{2000A&A...358..593S} Siess, L., Dufour, E., \&
Forestini, M.\ 2000, \aap, 358, 593

\bibitem[Simon et al.(2000)]{2000ApJ...545.1034S} Simon, M., Dutrey, A., 
\& Guilloteau, S.\ 2000, \apj, 545, 1034

\bibitem[Simon(2008)]{2008poii.conf..227S} Simon, M.\ 2008, The Power of 
Optical/IR Interferometry: Recent Scientific Results and 2nd Generation, 227

\bibitem[Steffen et al.(2001)]{2001AJ....122..997S} Steffen, A.~T., 
Mathieu, R.~D., Lattanzi, M.~G., et al.\ 2001, \aj, 122, 997

\bibitem[Tognelli et al.(2011)]{2011A&A...533A.109T} Tognelli, E., Prada
Moroni, P.~G., \& Degl'Innocenti, S.\ 2011, \aap, 533, A109

\bibitem[Torres et al.(2009)]{2009ApJ...698..242T} Torres, R.~M., Loinard, 
L., Mioduszewski, A.~J., \& Rodr{\'{\i}}guez, L.~F.\ 2009, \apj, 698, 242

\bibitem[Tuthill et al.(2000)]{2000PASP..112..555T} Tuthill, P.~G., 
Monnier, J.~D., Danchi, W.~C., Wishnow, E.~H., 
\& Haniff, C.~A.\ 2000, \pasp, 112, 555

\bibitem[Walter et al.(1988)]{1988AJ.....96..297W} Walter, F.~M., Brown, 
A., Mathieu, R.~D., Myers, P.~C., \& Vrba, F.~J.\ 1988, \aj, 96, 297

\bibitem[Wizinowich et al.(2000)]{2000SPIE.4007....2W} Wizinowich, P.~L., 
Acton, D.~S., Lai, O., et al.\ 2000, \procspie, 4007, 2

\bibitem[Wizinowich et al.(2004)]{2004SPIE.5491.1678W} Wizinowich, P.~L., 
Akeson, R.~L., Colavita, M.~M., et al.\ 2004, \procspie, 5491, 1678

\bibitem[Yelda et al.(2010)]{2010ApJ...725..331Y} Yelda, S., Lu, J.~R., 
Ghez, A.~M., et al.\ 2010, \apj, 725, 331

\end{thebibliography}
\end{document}